\documentstyle[aps]{revtex}

\begin{document}
\title{Rotational Viscosity in Linear Irreversible Thermodynamics and its
Application to Neutron Stars }

\author{Alfredo Sandoval-Villalbazo$^{1}$\thanks{Present address:
Relativity \& Cosmology Group, University of Portsmouth,
Portsmouth PO1 2EG, Britain. Email: alfredo.sandoval@port.ac.uk} ,
Ana Laura Garc\'{i}a-Perciante$^{1}$ and L.S Garc\'{i}a
Col\'{i}n$^{2}$\thanks{E-mail: lgcs@xanum.uam.mx}}
\address{1. Departamento de Ciencias, Universidad Iberoamericana,
Santa F\'{e}, Mexico D.F. Mexico\\ 2. Departamento de F\'{i}sica,
Universidad Aut\'{o}noma Metropolitana, Iztapala, Mexico D.F.
Mexico } \maketitle



\begin{abstract}
A generalized analysis of the local entropy production of a simple
fluid is used to show that, if intrinsic angular momentum is taken
into account, rotational viscosity must arise in the linear
non-equilibrium regime. As a consequence, the stress tensor of
dense rotating matter, such as the one present in neutron stars,
posseses a significant non-vansishing antisymmetrical part. A
simple argument suggests that, due to the extreme magnetic fields
present in neutron stars, the relaxation time associated to
rotational viscosity is large $\left( \approx 10^{21}\,s\right) $.
The formalism leads to generalized Navier-Stokes equations useful
in neutron star physics which involve vorticity in the linear
regime.
\end{abstract}

\section*{1. Introduction}

In a recent paper \cite{uno}, Rezania and Maartens have shown that relevant
effects due to vorticity are present in neutron star physics. Their
communication was primarily based in results derived from kinetic theory
which show that, if Boltzmann equation is treated by means of Grad's method,
a coupling between vorticity and shear necessarily arises \cite{dos}. This
paper is intended to present a study of an alternative role that vorticity
may play in neutron stars. This approach is based on a purely
phenomenological formalism capable of describing the thermodynamical role of
vorticity and rotational viscosity, without introducing any assumption
foreign to the concept of local equilibrium and the validity of the main
conservation laws .

A word of caution is here pertinent. In principle one could require the use
of relativistic irreversible thermodynamics to correctly cope with this
question as it has been developed in previous work \cite{tres}.
Nevertheless, in order to compare our results with those obtained in Ref.%
\cite{uno} we shall overlook this fact and perform a non-relativistic
analysis to emphasize on the role played by the rotational viscosity.

The paper is divided as follows: Section 2 gives a review of the
Meixner-Prigogine approach to non-equilibrium thermodynamics \cite{cuatro}.
Attention should be paid to the fact that in this scheme, nothing forbids
the existence of a non-vanishing antisymmetric part of the stress-energy
tensor. Section 3 is devoted to the analysis of the constitutive equations
relating the thermodynamical forces and fluxes, and its immediate effects in
neutron star dynamics. Section 4 includes a brief comparison between the
vorticity coupling suggested in Ref.\cite{uno} and the one here presented.
An outline of how it is possible to support the existence of a non-vanishing
antisymmetric part of the stress tensor with kinetic theory is also included.

\section*{2. Meixner-Prigogine Formalism Including Intrinsic Angular Momentum}

\subsection*{2.1 Conservation laws}

The Meixner-Prigogine approach to irreversible thermodynamics is based in
the consideration of the conservation laws of mass, linear momentum, angular
momentum and total energy, plus the assumption of local thermodynamical
equilibrium. The balance equations for mass and linear momentum, written in
differential form, are respectively given by:
\begin{equation}
\frac{\partial \,\rho }{\partial \,t}+\left( \rho \,u^{\ell }\right) _{;\ell
}=\frac{D\,\rho }{D\,t}+\rho \,\left( \,u^{\ell }\right) _{;\ell }=0
\label{uno}
\end{equation}
\begin{equation}
\frac{\partial \,\left( \rho \,u^{k}\right) }{\partial \,t}+\left( \rho
\,u^{\ell }\,u^{k}+\tau ^{\ell \,k}\right) _{;\ell }=\rho \frac{D\,u^{k}}{%
D\,t}+\,\tau _{;\ell }^{\ell \,k}=f^{k}  \label{dos}
\end{equation}
where the summation is taken for repeated indices, $\rho $ is the mass
density, $u^{\ell }$ the local hydrodynamic velocity, $\tau ^{\ell \,k}$ the
stress tensor and $f^{k}$ the external force density term. The forces of
interest in our formalism correspond to gravitational and magnetic fields.
Latin indices will run from $1$ to $3$, and $;$ indicates a covariant
derivative.

Total angular momentum per unit of mass in any local element is properly
described by an antisymmetric tensor $J^{k\ell \,}$. This tensor possesses
an orbital part $L^{\,k\ell }$ and an intrinsic contribution $S^{k\ell }$,
so that
\begin{equation}
J^{k\ell }=L^{k\ell }+S^{k\ell }  \label{a1}
\end{equation}
Since total momentum is conserved in purely rotating systems, we may write
\begin{equation}
\rho \frac{D\,J^{k\ell }}{D\,t}+\left( \,x^k\tau ^{a\,\ell }-x^\ell \tau
^{a\,k}\right) _{;\,a}=0  \label{a2}
\end{equation}
where $x^k$ stands for the position vector. It is important to stress at
this stage that the external force $f^k$ does not contribute to equation (%
\ref{a2}) for two reasons. The first obvious one is that the gravitational
force being radial generates no torques. The second one is a little bit more
subtle. Assuming that the external magnetic field $\overrightarrow{B}$ is
constant and points along the rotation axis, the Lorentz force generates
only a radial component whence again generates no torque. On the other hand
internal magnetic fields are accounted for in the intrinsic angular momentum
equation as it will be pointed out below. Now, orbital angular momentum per
unit of mass is given by
\begin{equation}
L^{\,k\ell }=x^ku^\ell -x^\ell u^k  \label{a25}
\end{equation}
From equations (\ref{dos}) and (\ref{a25}), it follows that, in the presence
of the fields of interest,
\begin{equation}
\rho \frac{D\,L^{k\ell }}{D\,t}+\left( \,x^k\tau ^{a\,\ell }-x^\ell \tau
^{a\,k}\right) _{;\,a}=\tau ^{k\ell \,}-\tau ^{\ell \,k\,}  \label{a3}
\end{equation}
so that, substracting equation (\ref{a3}) from equation (\ref{a2}) we have,
for the intrinsic angular momentum per unit of mass, that,
\begin{equation}
\rho \frac{D\,S^{k\ell }}{D\,t}=\tau ^{k\ell \,}-\tau ^{\ell \,k\,}\equiv
-2\tau ^{[k\ell ]}  \label{a35}
\end{equation}
where square brackets on indices indicate the skew part. Intrinsic angular
momentum, i.e. spin, in any local element of a neutron star\emph{\ is
assumed to be proportional to its local magnetization }$M^{k\ell }$, per
unit of area, per unit of charge, so that:
\begin{equation}
S^{k\ell }=\beta \,M^{k\ell }  \label{a4}
\end{equation}
where the coefficient $\beta $ has units of $(Lenght)^2$ and is
related to thermodynamical properties of the element. This
coefficient will be estimated in Section 4. It follows then, from
the corresponding balance of instrinsic rotational energy, per
unit of mass, $e_{\left[ rot\right] }$, that using equations
(\ref{a35}) and (\ref{a4})
\begin{equation}
\frac{\partial \,\,\left( \rho \,e_{\left[ rot\right] }\right)
}{\partial
\,t}+\left( \rho \,e_{\left[ rot\right] }u^k\right) _{;k}=\rho \frac{%
D\,e_{\left[ rot\right] }}{D\,t}=\frac \rho 4\frac{D\,\left(
\,\beta M_{k\ell }\,M^{k\ell }\right) }{D\,t}=-M_{k\ell }\tau
^{[k,\ell ]} \label{a5}
\end{equation}
where the second equality is due to the fact that, the rotational
contribution to the internal energy in a paramagnetic substance is
proportional to $B^{kl}B_{kl}$ and since the magnetization is
proportional to the magnetic field $B^{kl}$, equation (\ref{a5})
follows directly.

Let us now consider the energy balance. Since total energy is conserved, one
can write for this conservation law the expression,
\begin{equation}
\frac{\partial \,\left( \,e_{\left[ mec\right] }+e_{\left[ int\right]
}\right) }{\partial \,t}+\left( J_{\left[ mec\right] }^{k}+J_{\left[
Q\right] }^{k}+\,e_{\left[ int\right] }u^{k}\right) _{;k}=0  \label{seis}
\end{equation}
where $J_{\left[ Q\right] }^{k}$ is the heat flux density and $\,e_{\left[
int\right] }u^{k}$ the drift transport of internal energy. The explicit form
of the heat flux $J_{\left[ Q\right] }^{k}$ will be given later.

Multiplying equation (\ref{dos}) by the covariant vector $u_{\ell }$ and
making use of equations (\ref{uno}) and (\ref{a5}) we get that:
\begin{equation}
\frac{\partial \,\,e_{[mec]}}{\partial \,t}+\left(
J_{[mec]}^{k}\right) _{;k}=u_{\ell ;k}\,\tau ^{\ell \,k}-M_{k\ell
}\tau ^{\lbrack k,\ell ]} \label{tres}
\end{equation}
where
\begin{equation}
\,e_{\left[ mec\right] }=\frac 12\rho \,u^2+\rho \,\varphi +\rho
\,e_{\left[ rot\right] }  \label{cuatro}
\end{equation}
and
\begin{equation}
J_{\left[ mec\right] }^{k}=\frac{1}{2}\rho \,u^{2}u^{k}+\rho
\,\varphi \,u^{k}\,+\rho \,e_{\left[ rot\right] }u^{k}+u_{\ell
}\,\tau ^{\ell \,k} \label{cinco}
\end{equation}
where $\varphi $ stands for the gravitational potential and use has been
made of the fact that no dissipation arises from the magnetic field, since $%
u_{\ell }\,\varepsilon _{k\,n}^{\ell }\,u^{k}B^{n}=0$. If one now combines
equations (\ref{uno}) and (\ref{tres}), it is possible to obtain the balance
equation for the internal energy,
\begin{equation}
\,\frac{D\,e_{\left[ int\right] }}{D\,t}=-\left( J_{\,\left[ Q\right]
}^{\ell }\right) _{;\,\ell }-u_{\ell ;k}\,\tau ^{\ell \,k}+M_{\ell \,k}\tau
^{\lbrack \ell k]}  \label{siete}
\end{equation}
If the stress tensor is splitted into two parts, one related to the (scalar)
equilibrium pressure $p$ and one associated with the viscous pressure tensor
$\Pi ^{\ell \,k}$ , then $\tau ^{\ell \,k}=p\,\delta ^{\ell \,k}+\Pi ^{\ell
\,k},$ and equation (\ref{cinco}) reads:
\begin{equation}
\frac{D\,e_{\left[ int\right] }}{D\,t}=-\left( J_{\,\left[ Q\right] }^{\ell
}\right) _{;\,\ell }-u_{\ell ;k}\,\Pi ^{\ell \,k}-p\left(
u_{\,;a}^{a}\right) +M_{\ell \,k}\Pi ^{\lbrack \ell \,k]}  \label{ocho}
\end{equation}
It must be noticed that $\tau ^{\lbrack \ell \,,k]}=\Pi ^{\lbrack \ell \,k]}$%
.

\section*{3. Local Thermodynamical Equilibrium and Entropy Balance}

We now turn ourselves towards the question of introducing the entropy
production. For this purpose we invoke the local equilibrium assumption
which is crucial to the formulation of several versions of irreversible
thermodynamics \cite{ocho}. This assumption states that the local entropy $s$
of the system is a time-independent functional of the local scalar
thermodynamic densities, in this case the mass density $\rho \,\left(
x^{k},t\right) $ and the internal energy density $e_{\left[ int\right]
}\left( x^{k},t\right) $. Then,
\begin{equation}
s=s\left[ \rho \left( x^{a},t\right) ,e_{\left[ int\right] \,\,}\left(
x^{a},t\right) \right]  \label{s4}
\end{equation}
whence,
\begin{equation}
\frac{D\,s}{D\,t}=\left( \frac{\partial \,s}{\partial \,\rho }\right)
_{e_{\left[ int\right] \,}}\frac{D\rho }{D\,t}+\left( \frac{\partial \,s}{%
\partial \,e_{\left[ int\right] \,\,}}\right) _{\rho }\frac{De_{\left[
int\right] \,\,}}{D\,t}  \label{s45}
\end{equation}
By the local equilibrium assumption {\it per se}, the
thermodynamic coefficients are the local expressions of their
equilibrium counterparts,
\begin{equation}
\left( \frac{\partial \,s}{\partial \,\rho }\right) _{e_{\left[ int\right]
\,}}=-\frac{\,p}{\rho ^{2}T}\qquad \qquad \left( \frac{\partial \,s}{%
\partial \,e_{\left[ int\right] }}\right) _{\rho }=\frac{1}{T}  \label{s55}
\end{equation}
where $p$ and $T$ are the local pressure and temperature, respectively, and
the time rates of change of $\rho $ and $e_{\left[ int\right] \,\,}$ are
given through equations (\ref{uno}) and (\ref{siete}) respectively. Putting
all this information into equation (\ref{s45}) and after some algebraic
manipulation, one arrives at an expression of the form

\begin{equation}
_{\ }\frac{D\,s}{D\,t}+\left( \frac{J_{\,\left[ Q\right] }^{\ell }}{T}%
\right) _{;\,\ell }=-\frac{J_{\left[ Q\right] }^{\ell }}{T^{2}}\frac{%
\partial \,T}{\partial \,x^{\ell }}-\frac{1}{T}u_{\ell ;k}\,\Pi ^{\ell \,k}+%
\frac{1}{T}M_{\ell \,k}\Pi ^{\lbrack \ell \,k]}  \label{PE2}
\end{equation}
From standard irreversible thermodynamics, \emph{the entropy production }or
Clausius' uncompensated heat term $\Sigma $ is precisely the right hand side
of equation (\ref{PE2}) namely,
\begin{equation}
\Sigma =-\frac{J_{\left[ Q\right] }^{\ell }}{T^{2}}\frac{\partial \,T}{%
\partial \,x^{\ell }}-\frac{1}{T}u_{\ell ;k}\,\Pi ^{\ell \,k}+\frac{1}{T}%
M_{\ell \,k}\Pi ^{\lbrack \ell \,k]}  \label{PEnt2}
\end{equation}

Only when $\Sigma >0$, which depends on the nature of the constitutive
equations adopted, the theory will be consistent with the second law of
thermodynamics. We will come back to this questions in the next section.

\section*{4. Constitutive Relations and Neutron Star Dynamics}

\subsection*{4.1 Linear constitutive relations}

We are now able to propose constitutive relations for thermodynamical fluxes
and forces using Curie's principle [4, 5]. This principle states that, given
an expression such as (\ref{PEnt2}) for the entropy production in an
isotropic system, only tensors of equal rank may be coupled. According to
this idea, we must decompose tensors $\Pi ^{\ell \,k}$ and $u_{\ell ;k}$ in
their irreducible forms, namely:
\begin{equation}
\Pi ^{\ell \,k}=\Pi ^{[\ell \,k]}+\Pi ^{\left\langle \ell \,k\right\rangle
}+\Pi \,\delta ^{\ell \,k}  \label{diez}
\end{equation}
and
\begin{equation}
u_{\ell \,;\,k}=u_{\left[ \ell ;k\right] }+u_{\left\langle \ell
;\,k\right\rangle }+\left( u_{\,;a}^a\right) \,\delta _{\ell \,\,k}
\label{once}
\end{equation}
where angular brackets denote the symmetric trace-free part and $\Pi $
represents one third of the trace of $\Pi ^{\ell \,k}$. Introducing these
relations in equation (\ref{PEnt2}) we obtain:
\begin{equation}
\Sigma =-\frac{J_{\left[ Q\right] }^\ell }{T^2}\frac{\partial \,T}{\partial
\,x^\ell }-\frac 1T\Pi ^{[\ell \,k]}u_{\left[ \ell \,,\,k\right] }-\frac
1T\Pi ^{\left\langle \ell \,k\right\rangle }u_{\left\langle \ell
\,,\,k\right\rangle }-\frac 1T\Pi \,u_{\,;a}^a+\frac 1TM_{\ell \,k}\Pi
^{[\ell \,k]}  \label{doce}
\end{equation}
Since we wish to have $\Sigma >0$, linear constitutive relations are
admissible to relate thermodynamical fluxes and forces, so that
\begin{equation}
J_{[Q]}^\ell =-\kappa \,\frac{\partial \,T}{\partial \,x_\ell }
\label{Fourier1}
\end{equation}
\begin{equation}
\Pi =-\zeta \,u_{;a}^a  \label{Navier2}
\end{equation}
\begin{equation}
\Pi ^{\left\langle \ell \,k\right\rangle }=-2\eta \,u^{\left\langle \ell
;k\right\rangle }  \label{Navier3}
\end{equation}
and
\begin{equation}
\Pi ^{\left[ \ell \,k\right] }=-2\eta _R\,\,\left( u^{\left[ \ell
\,;k\right] }-M^{\ell \,k}\right)  \label{Navier4}
\end{equation}
The transport coefficients introduced here are $\kappa $, the heat
conductivity, $\eta ,$ the bulk viscosity, $\zeta $, the shear viscosity,
and $\eta _R$ \emph{the rotational viscosity}. The property sought for $%
\Sigma $ will be satisfied if these coefficients are themselves positive. As
is well known, this is borne out by experiment.

\subsection*{4.2 Navier-Stokes equations in the presence of rotational viscosity
and relaxation time.}

An immediate consequence of the validity of the constitutive relations (\ref
{Navier2}-\ref{Navier4}) is the effect of vorticity in the equation of
motion (\ref{dos}), since
\begin{equation}
\frac{\partial \,\tau ^{\ell \,k}}{\partial \,x^\ell }=\frac{\partial \,}{%
\partial \,x^\ell }\left[ \Pi ^{[\ell \,k]}+\Pi ^{\left\langle \ell
\,k\right\rangle }+\left( \Pi +p\right) \,\delta ^{\ell \,k}\right]
\label{doce.5}
\end{equation}
or
\begin{equation}
\frac{\partial \,\tau ^{\ell \,k}}{\partial \,x^\ell }=\frac{\partial \,}{%
\partial \,x^\ell }\left[ -2\eta _R\,\,\left( u^{\left[ \ell \,;k\right]
}-M^{\ell \,k}\right) -2\eta \,u^{\left\langle \ell \,;k\right\rangle
}-\zeta \left( \,u_{;a}^a\right) \delta ^{\ell \,k}\right] +\frac{\partial
\,p}{\partial \,x_k}  \label{trece}
\end{equation}
Now, as de Groot and Mazur observe \cite{cinco}, this effect of vorticity
involves a relaxation time obtainable from the assumption that $u^{\left[
\ell \,;k\right] }$ does not change substantially in a time interval $%
0<t<\Gamma $ where $\Gamma $ is the relaxation time characterizing the local
magnetization and defined below in equation (\ref{diecisiete}). Thus, from
equations (\ref{Navier4}) and (\ref{a35}), we have that,
\begin{equation}
2\Pi ^{\left[ \ell \,k\right] }=-4\eta _R\,\,\left( u^{\left[ \ell
;\,k\right] }-M^{\ell \,k}\right) =-\rho \frac{D\,S^{k\ell }}{D\,t}
\label{catorce}
\end{equation}
or, in terms of the local magnetization, equation (\ref{a4})
\begin{equation}
4\eta _R\,\,\left( u^{\left[ \ell ;\,k\right] }-M^{\ell \,k}\right) =\rho
\,\beta \frac{D\,\,M^{k\ell }}{D\,t}  \label{quince}
\end{equation}
which yields, under the uniformity assumption mentioned before, and assuming
a vanishing magnetization at $t=0$:
\begin{equation}
M^{\ell \,k}=u^{\left[ \ell ;\,k\right] }\left[
1-\exp\left({-\frac{4\eta _R}{\rho \,\beta }\,t}\right)\right]
\label{dieciseis}
\end{equation}
This coupling between intrinsic angular momentum and vorticity involves a
relaxation time given by:
\begin{equation}
\Gamma =\frac{\rho \,\beta }{4\eta _R}  \label{diecisiete}
\end{equation}
Densities in neutron stars are extremely large, typically near to
$10^{18}$ kg m$^3$, so that $\Gamma $ essentially depends on the
ratio $\beta /\eta _R$. Its magnitude will be explored below.

\subsection*{4.3 Order of magnitude for the relaxation time}

The other parameter left in order to obtain a good idea about the
relevance of this vorticity-magnetization coupling is ${\beta
}/{\eta _{R}}$ in
equation (\ref{diecisiete}). Strong magnetic fields in the star, close to $%
B=10^{6}$ Tesla, totally associated with local spin, suggest that
the coefficient $\beta $ may be estimated by a very simple
argument, namely the relation between local magnetization, per
unit of area, per unit of charge, and spin per kilogram of mass of
a perfect paramagnetic solid, which is given by \cite{seis}:
\begin{equation}
S^{k\ell }=\frac{1}{2\pi }\frac{h\,}{\mu \,\,{\cal L}(\mu B/kT
)}\,M^{k\ell }  \label{dieciocho}
\end{equation}
where $h$ is Planck's constant, $\mu =10^{-24}$ A.m$^{2}$ is
Bohr's magneton and ${\cal L}(T)$ is the Langevin function at
temperature $T.$ In our approximation, ${\mu \,B}/{k\,T}\approx
10^{-5}$. Direct comparison with equation (\ref{a4}) yields,
\begin{equation}
\beta =\frac{1}{2\pi }\frac{h\,}{\mu \,\,{\cal L}({\mu B/kT})}%
\approx 10^{-6}\,{\rm m}^{2}  \label{diecinueve}
\end{equation}
and, thus
\begin{equation}
\Gamma \approx \frac{10^{11}{\rm Pa. s}^{2}}{\eta _{R}}
\label{veinte}
\end{equation}
which may be quite significant if $\eta _{R}$\ is a relatively
small number. Estimates of this coefficient are scarce in the
literature, but even if we assume that it is of the order of
$10^{3}$ Pa s$^{2}$, $\Gamma $ would be very large, if compared
with relaxation times of the so-called ``fast
variables'' associated with extended irreversible thermodynamics \cite{siete}%
.

\section*{5. Final Remarks}

It must be stated that the formalism here developed is not in conflict with
the one presented in Ref.[1], where vorticity appeared without taking into
account intrinsic spin, but by means of a coupling between \emph{shear} and
vorticity. Whether the vorticity effects associated to intrinsic spin are
relevant or not to stability of neutron stars is not clear. Boltzmann
equation itself cannot predict values of rotational viscosity, since it
leads to symmetric stress tensors and involves no information about internal
degrees of freedom. A kinetic treatment involving the generalization
suggested by Waldmann and Snider [9, 10] may be useful in order to compute
precisely the value of $\eta _{R}$ in equation (\ref{diecisiete}) and so
conclude about the real significance of this relaxation time.

On the other hand, if the dynamical effect of spin were negligible in dense
systems like neutron stars, then, according to equation (\ref{diecisiete}),
rotational viscosity must have large values and, from this point of view,
further studies related to astrophysical implications of this coefficient
should developed.

\section*{Acknowledgements}

The authors wish to thank Roy Maartens for valuable comments and suggestions
that have improved the quality this work.


\end{document}